\documentclass[a4paper,aps,pra,twocolumn,nofootinbib]{revtex4-1}

\usepackage{amsmath}
\usepackage{amsfonts}
\usepackage{amssymb}

\usepackage{graphicx}

\usepackage{bm}

\newcommand{\textindex}[1]{\mbox{\tiny #1}}

\begin{document}
\title{Nonlinear transformation optics and engineering of the Kerr effect}
\author{L.~Bergamin}
\affiliation{KB\&P GmbH, Fliederweg 10, 3007 Bern, Switzerland}
\email{Luzi.Bergamin@kbp.ch}
\author{P.~Alitalo and S.A.~Tretyakov}
\affiliation{Department of Radio Science and Engineering, Aalto University, Otakaari 5A, 00076 Aalto, Finland}
\date{December 23, 2011}

\begin{abstract}
The concept of transformation optics is extended to nonlinear electrodynamics. It is shown that transformation optics favors implicit constitutive relations in terms of energy densities $\bm D \cdot \bm E$ and $\bm B \cdot \bm H$ rather than $\bm E^2$ and $\bm H^2$. The Kerr nonlinearity is studied in detail and the  transformation optics based engineering of self-interaction effects is discussed. As a specific example we introduce transformation optics applied on a self-focusing field configuration.
\end{abstract}


\maketitle

\section{Introduction}
Transformation optics (TO) \cite{Pendry:2006Sc,Leonhardt:2006Sc,Leonhardt:2006Nj} has become one of the most powerful tools in the design of artificial media (metamaterials). TO allows the derivation of a linear constitutive relation directly from a desired trajectory of light -- in a completely algebraic way. The key ingredient of TO is to define the trajectory in the specified medium as a result of a transformation applied on the trajectories in free space. Unlike normal coordinate transformations, under which physics is invariant, the field lines are considered to be attached to the coordinates during the transformation, thereby changing the physics and deriving the constitutive law of the desired medium. Despite some important differences, which are explained in Sect.\ \ref{se:II}, the characteristics of transformation media remain very close to those of the transformed empty space. Thus, transformation media are often interpreted as media that mimic (a transformation of) empty space. This possibility greatly stimulated research in metamaterials: besides the original proposal of a cloak \cite{Pendry:2006Sc,Leonhardt:2006Sc} (a point is blown up to a sphere), the large number of proposals include beam shifters and benders \cite{Rahm:2008Pl}, field concentrators \cite{Rahm:2008Cn} (a large cylinder is compressed into a smaller one), flat hyperlenses by means of stretching coordinates \cite{Gaillot:2008Nj} and, more recently, a carpet cloak \cite{Li:2008Cc}. Also, attempts have been made to generalize the concept of TO: instead of a Euclidean empty space a curved space can be mimicked \cite{Leonhardt:2009Ec}, which can be used to reproduce effects known in general relativity \cite{Genov:2009Np}, and more complicated geometric \cite{Bergamin:2008Pa} or even direct field transformations \cite{Tretyakov:2008Gf} can replace the simple coordinate transformation. It should be mentioned that TO also stimulated research outside the field of electromagnetics, notably similar principles can be applied to matter waves in quantum mechanics \cite{Zhang:2008Qc} or to sound waves in acoustics \cite{Cummer:2007Ac,Chen:2007Ac}. An extensive review of the mathematical principles of TO can be found in Ref.\ \cite{Leonhardt:2008Oe}, a recent summary of various applications has been presented in Ref.\ \cite{Chen:2010Nm}.

Since transformation media are linked to empty space via a coordinate transformation it follows immediately that all transformation media must be linear. Transformation of time in standard TO leads, in general, to transformation media with time dependent media parameters. In these media effects such as frequency conversion are possible \cite{Cummer:2011Jo}, although the material is linear. Similar conclusions apply to the aforementioned extensions except for the field transforming metamaterials \cite{Tretyakov:2008Gf}, which however lack a simple geometric interpretation. Moreover, it was shown recently that the constitutive relation obtained from TO is unique up to some rescaling factors \cite{Favaro:2010Nb}. In this paper the idea of TO is applied to nonlinear electrodynamics. From the conclusions drawn above the basic strategy of this task is immediate: if a nonlinear medium is to be obtained after the transformation, one has to start with nonlinearities. Therefore, a nonlinear medium cannot be seen as a medium ``mimicking free space'' as is the case in standard TO. Still, the transformational approach can provide a tool with which solutions in complex, nonlinear media can be derived from much simpler -- though nonlinear -- situations. The very generic setup of TO guarantees that the solutions of the Maxwell equations in the complex material follow algebraically from the solutions in the simpler situation and -- as in linear TO -- this algebraic transformation has an immediate geometric interpretation. In this way, whole classes of nonlinear media can be related via TO to one particular, simple case (as, for example, the Kerr nonlinearity). Due to symmetry properties it will be derived that the nonlinear interactions of these classes are described in terms of energy densities $\bm D \cdot \bm E$ and $\bm B \cdot \bm H$ rather than $\bm E^2$ and $\bm H^2$.

Most applications of linear TO are supposed to be independent of the electromagnetic field configuration -- as an example, the TO cloak \cite{Pendry:2006Sc,Leonhardt:2006Sc} theoretically produces an invisible area for any incoming electromagnetic field. All these applications straightforwardly extend to nonlinear TO. However, even simple nonlinear media such as a Kerr material exhibit interesting self-interaction effects, which depend not just on the medium parameters but also on the details of the electromagnetic fields propagating in the nonlinear medium. As is shown, such effects can also be engineered effectively using nonlinear TO. Of course, the workability of the resulting devices will be restricted to exactly those field configurations that exhibit the interaction effects already in the original untransformed medium, but the main advantages of TO (the geometrically intuitive construction and the algebraic derivation of the solutions) are completely retained.

\section{Transformation optics for linear media}
\label{se:II}
In its standard formulation, TO is restricted to a special class of linear media that may be given the interpretation of mimicking the free space solutions. To facilitate the discussion of specific issues appearing with nonlinear TO, this simpler concept is introduced in this section -- in doing so, some technical aspects important for this paper are highlighted.

Since TO deals with generic, curvilinear coordinates, a suitable notation taking care of the covariance properties in such coordinates should be used. Here, we use component notation in conjunction with the Einstein summation convention. The Maxwell equations in this notation read
\begin{align}
\label{maxwell}
 \nabla_i B^i &= 0\ , & \frac{d}{d t} B^i + \epsilon^{ijk}\partial_j E_k &= 0\ , \\
\label{maxwell2}
 \nabla_i D^i &= \rho\ , & \epsilon^{ijk} \partial_j H_k - \frac{d}{d t}  D^i &= j^i\ .
\end{align}
Here, $\partial_i = \partial/\partial x^i$ is the directional and $\nabla_i$ is the covariant derivative with respect to the spatial metric metric $\gamma_{ij}$ and its determinant $\gamma$:
\begin{equation}
\label{covder}
 \nabla_i A^i = (\partial_i + \Gamma^j_{j i}) A^i = \frac{1}{\sqrt{\gamma}} \partial_i(\sqrt{\gamma}A^i)\ .
\end{equation}
Finally, $\epsilon^{ijk}$ is the Levi-Civita tensor,
\begin{equation}
\begin{split}
 \epsilon^{ijk} &= \frac{1}{\sqrt{\gamma}} [ijk]\ ; \\ [ijk] &= \begin{cases} 1 & \text{$(ijk)$ are an even permutation of (123),}\\ -1 & \text{$(ijk)$ are an odd permutation of (123),} \\ 0 & \text{otherwise.} \end{cases}
\label{eq:levicivita}
\end{split}
\end{equation}
Standard TO starts with the Maxwell equations complemented by the free space constitutive relations
\begin{align}
\label{vacrel}
 D^i &= \varepsilon_0 \gamma^{ij} E_j\ , & B^i &= \mu_0 \gamma^{ij} H_j\ .
\end{align}
In free space the trajectories of light are straight lines, and to obtain a transformation medium a coordinate transformation is applied to this situation. Of course, a normal coordinate transformation does not affect physics, but in TO the electromagnetic field lines are considered as ``attached'' to the coordinates, such that the coordinate transformation does not represent an invariance but in fact deforms and stretches the trajectories of light. The resulting transformation medium is thus not physically equivalent to the original free space situation, but nonetheless the complete solution of the Maxwell equations in the medium is obtained in an algebraic way from the free space solution.

To understand the technical aspects of TO, the covariance properties of all quantities involved must be understood. In index notation, these covariance properties are encoded in the position of the indices. $\bm D = [D^i]$ and $\bm B = [B^i]$ with upper indices are vectors (column vectors), while $\bm E = [E_i]$ and $\bm H = [H_i]$ are co-vectors (row vectors). By means of the scalar product a column and a row vector are mapped onto a scalar, e.g.\ $\bm D \cdot \bm E = D^i E_i$, but no similar procedure exists for two vectors or two co-vectors; in terms of the index notation this manifests in the rule that summation is only allowed over opposite index positions. Vectors are transformed into co-vectors by means of the metric, $D_i = D^j \gamma_{ji}$, co-vectors into vectors by means of the inverse metric, $E^i = \gamma^{ij} E_j$. This manipulation allows to define a scalar quantity $\bm E^2 = E_i \gamma^{ij} E_j$ with an analogous expression for vectors. Since the metric in Cartesian coordinates is simply the unit matrix, no essential difference between vectors and co-vectors exists in that case, but in generic coordinates a careful distinction of the two kinds of objects is indispensable. This is seen when considering coordinate transformations, the first step in the TO program.

Under a transformation $\bm x \rightarrow \bar{\bm x}(x)$ with the Jacobian
\begin{equation}
\label{jacobian}
 T^{i}{}_{j} = \frac{\partial \bar x^i}{\partial x^j}
\end{equation}
each \emph{index} transforms linearly, again following the rule of summation over opposite indices and the obvious requirement that a summation can only be performed over indices that belong to the same coordinate system. Therefore,\footnote{In this work we consider exclusively proper coordinate transformations, $\det[T^{i}{}_{j}]>0$. The extension of TO to include improper coordinate transformations includes some technical difficulties, but does not yield any new types of material \cite{Bergamin:2010Nr}.}
\begin{align}
\label{fieldtrafo}
 \bar D^i \left(\bar x(x)\right) &= \frac{\partial \bar x^i}{\partial x^j} D^j(x)\ , \\ \bar E_i \left(\bar x(x)\right) &= \frac{\partial x^j}{\partial \bar x^i} E_j(x)\ , \\ \bar \gamma_{ij}\left(\bar x(x)\right) &= \frac{\partial x^k}{\partial \bar x^i} \gamma_{kl}(x) \frac{\partial x^l}{\partial \bar x^j}\ .
\end{align}
To express the effect of such a transformation on Eqs.\ \eqref{maxwell}, \eqref{maxwell2} and \eqref{vacrel} we can merely put a bar over each quantity, since physical processes are independent of the choice of coordinates. The content of the equations remains exactly the same. As mentioned above, TO makes use of special transformations, where the field lines of electromagnetics are considered to be stitched to the coordinates. More technically this means: While the barred metric should be used in the constitutive law \eqref{vacrel}, the Maxwell equations should be written in terms of the original metric without a bar. This ensures that the solutions of the Maxwell equations indeed are written in the coordinates $\bm x$ with metric $\gamma_{ij}$, but light follows straight lines with respect to the metric $\bar \gamma_{ij}$ and thus behaves ``as if'' the metric were $\bar \gamma_{ij}$. This manipulation is possible thanks to the fact that the metric appears in the Maxwell equations only via the covariant derivative \eqref{covder} and therein the metric $\bar \gamma_{ij}$ can easily be changed back to the original metric $\gamma_{ij}$ by rescalings \cite{Leonhardt:2006Nj,Bergamin:2010Nr}
\begin{align}
\label{rescaling}
 \tilde D^i &= \frac{\sqrt{\bar \gamma}}{\sqrt{\gamma}} \bar D^i\ , & \tilde B^i &= \frac{\sqrt{\bar \gamma}}{\sqrt{\gamma}} \bar B^i\ , & \tilde E_i &= \bar E_i\ , & \tilde H_i &= \bar H_i\ ,
\end{align}
which modify the transformed vacuum relation to the constitutive law of TO
\begin{align}
\label{TOrel}
 \tilde D^i &= \frac{\sqrt{\bar \gamma}}{\sqrt{\gamma}} \varepsilon_0 \bar \gamma^{ij} \tilde E_j\ , & \tilde B^i &= \frac{\sqrt{\bar \gamma}}{\sqrt{\gamma}} \mu_0 \bar \gamma^{ij} \tilde H_j\ .
\end{align}
It should be stressed again that these equations have to be understood as equations in the original coordinates $\bm x$.

The rescaling procedure \eqref{rescaling} is not a symmetry of electrodynamics since it maps a free space solution of the Maxwell equations onto a solution in a certain linear medium. Therefore, there must exist some physical differences between the free space and the TO solution. Again, the component notation proves extremely powerful for clarifying some of these differences. Obviously, in any free space solution $\bm D$ is parallel to $\bm E$, if the latter is transformed into a vector:
\begin{align}
 (D^1,D^2,D^3) &\parallel (E^1,E^2,E^3) = (\gamma^{1i} E_i,\gamma^{2i} E_i,\gamma^{3i}E_i)\ , \\(\bar D^1,\bar D^2,\bar D^3) &\parallel (\bar E^1,\bar E^2,\bar E^3) = (\bar \gamma^{1i} \bar E_i,\bar \gamma^{2i} \bar E_i,\bar \gamma^{3i} \bar E_i)\ .
\end{align}
Evidently, being parallel is a coordinate independent statement and thus holds in the transformed coordinate system $\bar{\bm x}$ as well. The situation is different in TO: Here the constitutive law is written in terms of the matrix $\bar \gamma_{ij}$, the metric in a fictuous ``electromagnetic space.'' Still, co-vectors are transformed into vectors by means of the true metric of the physical space, which is $\gamma_{ij}$. Therefore,
\begin{multline}
 (\tilde D^1,\tilde D^2,\tilde D^3) = \frac{\sqrt{\bar \gamma}}{\sqrt{\gamma}} \varepsilon_0 (\bar \gamma^{1j} \tilde E_j,\bar \gamma^{2j} \tilde E_j,\bar \gamma^{3j} \tilde E_j ) \\ \nparallel (\gamma^{1i} \tilde E_i,\gamma^{2i} \tilde E_i,\gamma^{3i} \tilde E_i)\ .
\end{multline}
Thus, in a transformation medium $\bm D$ and $\bm E$ fail to be parallel in general, an observation that has important consequences in what follows. This observation can be generalized: Quantities (except the constitutive relation) that depend on the metric are not mapped correctly within TO and, thus, represent a physical difference between the transformation medium and the original free space solution\footnote{The metric dependence of Gauss's law and its magnetic counterpart is somewhat accidental and easily can be removed in the so-called premetric formulation of electrodynamics \cite{Hehl:2003}. In the present formulation this dependence can be removed by means of rescalings \eqref{rescaling} since the covariant derivative just depends on the determinant of the metric, but not on its individual components.}. Besides the relative orientation of vectors and co-vectors discussed above this applies in particular to the stress-energy-momentum tensor \cite{Bergamin:2008Pa} and the Poynting vector as a part of the latter.  

It should be mentioned that more rigorous, coordinate independent formulations of TO are possible. Also, since electrodynamics is inherently relativistic, transformations mixing space and time can be addressed, which yields an extension of TO to bi-anisotropic media \cite{Leonhardt:2006Nj}. However, the interpretation of such transformations is less immediate in the presence of nonlinear media, thus we will restrict ourselves to purely spatial transformations in this work.

\section{Transformation optics for nonlinear media: a generic ansatz}

\label{sec:AGenericAnsatz}

As mentioned above, the only strategy for obtaining a nonlinear transformation medium is to choose as a starting point nonlinear medium constitutive relations instead of the free space relations \eqref{vacrel}. In this section we analyze how the general ansatz of a nonlinear medium
\begin{align}
\label{nonlin1A}
 D^i &= \left(\varepsilon^{ij} + \chi_{1}^{ijk} E_k + \chi_2^{ijkl} E_k E_l + \ldots\right) E_j\ , \\
 \label{nonlin1B}
 B^i &= \left(\mu^{ij} + \xi_{1}^{ijk} H_k + \xi_2^{ijkl} H_k H_l + \ldots\right) H_j\ ,
\end{align}
behaves in TO. Here, $\varepsilon^{ij} = \varepsilon_0 \varepsilon_R^{ij}$ and $\mu^{ij} = \mu_0 \mu_R^{ij}$ are the -- in principal arbitrary -- linear permittivity and permeability tensors, while the tensors $\chi_1^{ijk}$, $\chi_2^{ijkl}$ etc.\ and $\xi_1^{ijk}$, $\xi_2^{ijkl}$ etc.\ parametrize the nonlinear contributions. A transformational approach to design new media starting from Eqs.\ \eqref{nonlin1A} and \eqref{nonlin1B} is straightforward. In the first step we apply the coordinate transformation; the nonlinear constitutive law is invariant if all medium parameters transform as tensors:
\begin{gather}
\label{chitrans}
 \bar \varepsilon^{i'j'} = T^{i'}{}_{i} T^{j'}{}_{j} \varepsilon^{ij} \qquad  \bar \chi_{1}^{i'j'k'} = T^{i'}{}_{i} T^{j'}{}_{j} T^{k'}{}_{k} \chi_{1}^{ijk} \\ \bar \chi_{2}^{i'j'k'l'} = T^{i'}{}_{j} T^{j'}{}_{j} T^{k'}{}_{k} T^{l'}{}_{l} \chi_{2}^{ijkl} \\
 \label{xitrans}
  \bar \mu^{i'j'} = T^{i'}{}_{i} T^{j'}{}_{j} \mu^{ij} \qquad  \bar \xi_{1}^{i'j'k'} = T^{i'}{}_{i} T^{j'}{}_{j} T^{k'}{}_{k} \xi_{1}^{ijk} \\ \bar \xi_{2}^{i'j'k'l'} = T^{i'}{}_{j} T^{j'}{}_{j} T^{k'}{}_{k} T^{l'}{}_{l} \xi_{2}^{ijkl} 
\end{gather}
In the second step the rescaling rules \eqref{rescaling} are applied, which do not affect the new nonlinear contributions at all. Thus, the result
\begin{align}
\label{GenNonLinTrans}
 \tilde D^i &= \frac{\sqrt{\bar \gamma}}{\sqrt{\gamma}} \left(\bar \varepsilon^{ij} + \bar \chi_{1}^{ijk} \tilde E_k + \bar \chi_2^{ijkl} \tilde E_k \tilde E_l + \ldots\right) \tilde E_j\ , \\
\label{GenNonLinTrans2}
  \tilde B^i &= \frac{\sqrt{\bar \gamma}}{\sqrt{\gamma}} \left(\bar \mu^{ij} + \bar \xi_{1}^{ijk} \tilde H_k + \bar \xi_2^{ijkl} \tilde H_k \tilde H_l + \ldots\right) \tilde H_j\ ,
\end{align}
is immediate.

This result shows that TO extends straightforwardly to nonlinear electrodynamics. Nonetheless, the interpretation of the result is not as simple as in standard TO, but it can be cast into the following statement: The transformation medium \eqref{GenNonLinTrans} and \eqref{GenNonLinTrans2} behaves as if it were the original medium \eqref{nonlin1A} written in coordinates $\bar{\bm x}$, but in fact it is the physically different medium \eqref{GenNonLinTrans}/\eqref{GenNonLinTrans2} in coordinates $\bm x$. Here, ``behaves as'' has the same meaning as in standard TO, namely the solution of the Maxwell equation in the transformation medium with coordinates $\bm x$ is -- up to rescalings \eqref{rescaling} -- equivalent to the solution of the original medium in coordinates $\bar{\bm x}$, but this analogy does not extend to any quantity that depends on the metric. In this way it is possible to design complicated nonlinear media starting from a simple, though still nonlinear, situation.

It was mentioned above that four-dimensional spacetime transformations instead of purely spatial coordinate transformations can be applied in standard TO \cite{Leonhardt:2006Nj}. Since transformations mixing space and time transform electric into magnetic fields and vice versa, such transformations are associated with bi-anisotropic media. In the presence of nonlinearities, the mixing of electric and magnetic fields affects not just the linear term, but also the nonlinear expansion. Thus, such transformations would result in nonlinear electric-magnetic mixing terms, which might be difficult to interpret.

\section{Expansion in energy densities and the Kerr nonlinearity}

\subsection{The general expansion}
\label{sec:ExpansionInEnergyDensities}

In many applications of nonlinear electromagnetics the nonlinearities entering the constitutive law, the polarization density or derived quantities such as the refractive index are parametrized in terms of $\bm E^2$ and $\bm H^2$. Instead of the generic law \eqref{nonlin1A} and \eqref{nonlin1B} this suggests the common ansatz
\begin{align}
\label{nonlin1.1}
 D^i &= \left(\varepsilon^{ij} + \chi_{2}^{ij} \bm E^2 +  \chi_{4}^{ij} (\bm E^2)^2 + \ldots\right) E_j\ , \\ \label{nonlin1.1A} B^i &= \left(\mu^{ij} + \xi_{2}^{ij} \bm H^2 +  \xi_{4}^{ij} (\bm H^2)^2 + \ldots\right) H_j \ ,
\end{align}
where the nonlinear terms are now parametrized in terms of sequences of matrices $\chi_{I}^{ij}$ and $\xi_{I}^{ij}$ ($I=2,4,6,\ldots$). Of course, we could try to use the standard formulae of TO to derive the constitutive law of transformation media starting with a special case of Eqs.\ \eqref{nonlin1.1} and \eqref{nonlin1.1A}. But at this point, we should notice that the expansion parameters in Eqs.\ \eqref{nonlin1.1} and \eqref{nonlin1.1A},
\begin{align}
 \bm E^2 &= E_i \gamma^{ij} E_j\ , & \bm B^2 &= H_i \gamma^{ij} H_j\ ,
\label{eq:E2B2parameters}
\end{align}
implicitly depend on the metric. From the transformation rule
\begin{multline}
\label{scalarinv}
 \bm E^2 = E_i \gamma^{ij} E_j \Longrightarrow \bar{\bm E^2} = \bar E_i \bar \gamma^{ij} \bar E_j\\ = (\bm E \cdot T^{-1})_i (T\cdot\gamma\cdot T^T)^{ij} (T^{-1T}\cdot \bm E)_j \equiv \bm E^2\ ,
\end{multline}
we thus would find that after the transformation the resulting nonlinear medium,
\begin{align}
\label{nonlin2}
 \tilde D^i &= \frac{\sqrt{\bar \gamma}}{\sqrt{\gamma}} \left(\bar \varepsilon^{ij} + \bar \chi_2^{ij }\bm E^2 + \chi_{4}^{ij} (\bm E^2)^2  + \ldots\right) \tilde E_j\ , \\ \label{nonlin2A} \tilde B^i &= \frac{\sqrt{\bar \gamma}}{\sqrt{\gamma}} \left(\bar \mu^{ij} + \bar \xi_{2}^{ij} \bm H^2 + \bar \xi_{4}^{ij} (\bm H^2)^2 + \ldots\right) \tilde H_j \ ,
\end{align}
is not expanded in terms of the fields in the transformation medium $\tilde{\bm E}^2 = \tilde E_i \gamma^{ij} \tilde E_j$, $\tilde{\bm H}^2 = \tilde H_i \gamma^{ij} \tilde H_j$, but in terms of the fields in the \emph{original} medium as in Eqs.\ \eqref{nonlin1.1} and \eqref{nonlin1.1A}.

This difficulty can be overcome by replacing the conventional expressions \eqref{nonlin1.1} and \eqref{nonlin1.1A} with equivalent metric-independent expressions. To this end, we notice that in an arbitrary (and potentially anisotropic) material the energy densities are not proportional to $\bm E^2$ and $\bm H^2$. Instead, an invariant energy density of the electromagnetic field is given by
\begin{equation}
U = \bm D \cdot \bm E + \bm B \cdot \bm H
\label{eq:energydensity}
\end{equation}
which suggests using the electric and magnetic energy densities,
\begin{align}
U_E &= \bm D \cdot \bm E = D^i E_i\ , & U_M &= \bm B \cdot \bm H = B^i H_i\ ,
\label{eq:EBenergydensities}
\end{align}
as expansion parameters. As shown by these equations, this approach automatically removes all concerns about metric dependence since the invariant energy densities are indeed independent of any spatial metric. This means that due to both physical and symmetry arguments one should start with the implicit constitutive law
\begin{align}
\label{nonlin4A}
 D^i &= \left(\varepsilon^{ij} + \hat \chi_{2}^{ij} \bm D \cdot \bm E +  \hat \chi_{4}^{ij} (\bm D \cdot \bm E)^2 + \ldots\right) E_j\ , \\
\label{nonlin4B}
 B^i &= \left(\mu^{ij} + \hat \xi_{2}^{ij} \bm B \cdot \bm H +  \hat \xi_{4}^{ij} (\bm B \cdot \bm H)^2 + \ldots\right) H_j \ .
\end{align} 
After this modification, all formulae of standard TO again can be used straightforwardly to arrive at the transformation medium
\begin{align}
\label{nonlin6A}
 \tilde D^i &=  \left(\frac{\sqrt{\bar \gamma}}{\sqrt{\gamma}} \bar \varepsilon^{ij} + \hat{\bar \chi}_{2}^{ij} \tilde{\bm D} \cdot \tilde{\bm E} +  \frac{\sqrt{\gamma}}{\sqrt{\bar \gamma}} \hat{\bar \chi}_{4}^{ij} (\tilde{\bm D} \cdot \tilde{\bm E})^2 + \ldots\right) \tilde E_j\ ,\\
 \label{nonlin6B}
 \tilde B^i &=  \left(\frac{\sqrt{\bar \gamma}}{\sqrt{\gamma}} \bar \varepsilon^{ij} + \hat{\bar \xi}_{2}^{ij} \tilde{\bm B} \cdot \tilde{\bm H} +  \frac{\sqrt{\gamma}}{\sqrt{\bar \gamma}} \hat{\bar \xi}_{4}^{ij} (\tilde{\bm B} \cdot \tilde{\bm H})^2 + \ldots\right) \tilde H_j\ .
\end{align}
These equations represent the simplified version of Eqs.\ \eqref{GenNonLinTrans} and \eqref{GenNonLinTrans2} in the case where the nonlinear response can be expanded in terms of the energy densities rather than the electric and magnetic field monomials.

It should be realized that the models \eqref{nonlin1.1}/\eqref{nonlin1.1A} and \eqref{nonlin4A}/\eqref{nonlin4B} in general are two different limits of the the generic expansion \eqref{nonlin1A}/\eqref{nonlin1B}. To illustrate this fact let us assume a specific situation where cubic and higher nonlinearities can be neglected. Clearly, the model \eqref{nonlin1.1}/\eqref{nonlin1.1A} corresponds to the situation, where the third and fourth order tensors in \eqref{nonlin1A}/\eqref{nonlin1B} may be written as
\begin{align}
\chi_1^{ijk}&=\xi_1^{ijk}=0\ , & \chi_2^{ijkl}&= \chi_2^{ij}\gamma^{kl}\ , & \xi_2^{ijkl}&= \xi_2^{ij}\gamma^{kl}\ .
\end{align}
Conversely, \eqref{nonlin1A}/\eqref{nonlin1B} correspond to the limit
\begin{align}
\chi_1^{ijk}&=\xi_1^{ijk}=0\ , & \chi_2^{ijkl}&= \hat \chi_2^{ij}\varepsilon^{kl}\ , & \xi_2^{ijkl}&= \hat\xi_2^{ij}\mu^{kl}\ .
\end{align}
Clearly the two descriptions are equivalent if and only if $\gamma^{ij}\propto \varepsilon^{ij}\propto \mu^{ij}$, in particular in the case of the Kerr nonlinearity discussed below.

\subsection{The Kerr nonlinearity}
\label{sec:TheKerrNonLinearity}

The theory in the previous subsection encompasses large classes of nonlinear media. In this subsection a further restriction to the special case of the Kerr nonlinearity is made. Kerr media are described by the constitutive law\footnote{Evidently, in the special case of the Kerr medium the electric energy density is proportional to $\bm E^2$, which justifies the standard approach. However, for anisotropic media this ansatz remains problematic.}
\begin{align}
D^i &= \left( \varepsilon + \chi \bm E^2\right) \gamma^{ij} E_j\ , & B^i &= \mu_0 \gamma^{ij} H_j\ ,
\label{eq:Kerr1}
\end{align}
which, according to the discussion in the previous section, should be rewritten as
\begin{align}
D^i &= \left( \varepsilon + \hat \chi \bm D \cdot \bm E \right) \gamma^{ij} E_j\ , & B^i &= \mu_0 \gamma^{ij} H_j\ ,
\label{eq:Kerr2}
\end{align}
with
\begin{equation}
\hat \chi = \frac{\chi}{\varepsilon} + \mbox{higher-order contributions,}
\label{eq:HatChi}
\end{equation}
and it is assumed that quartic and higher-order contributions to the nonlinearity term are irrelevant. Taking the standard Kerr medium as a starting point, a coordinate transformation $\bm x \Rightarrow \bar{\bm x}$ yields a transformation medium
\begin{align}
\tilde D^i &=  \left( \frac{\sqrt{\bar \gamma}}{\sqrt{\gamma}} \varepsilon + \hat \chi \tilde{\bm D} \cdot \tilde{\bm E} \right) \bar \gamma^{ij} \tilde{E}_j\ ,  & \tilde B^i &= \mu_0 \frac{\sqrt{\bar \gamma}}{\sqrt{\gamma}}  \bar \gamma^{ij} \tilde H_j\ .
\label{eq:kerr4}
\end{align}
Besides the transformation of the nonlinearity itself it is important to realize that the resulting medium has a non-trivial, though linear, magnetic response. 

Instead of the constitutive law \eqref{eq:Kerr1}, implications of the optical Kerr effect are often discussed in terms of an intensity dependent refractive index profile
\begin{align}
 n &= n_0 + n_2 I\ , & n_0 &= \sqrt{\varepsilon_R}\ , \\ n_2 &= \frac{12 \pi^2}{n_0^2 c} \chi\ , &  I &= \frac{n_0 c}{8\pi} \bm{\mathcal E}^2\ ,
\label{eq:kerr5}
\end{align}
where $\bm{\mathcal E}$ is the Fourier component of a monochromatic wave $\bm E(t)$. In terms of this relation several self-interaction effects can be described, see e.g.\ Ref.\ \cite{Boyd:2008}. As a specific example, let us briefly summarize the effect of self-focusing, whose behavior in TO is studied below. For powers of the incident beam that are greater than the critical power
\begin{equation}
P_{\textindex{cr}} = \frac{0.372 \pi \lambda_0^2}{8 n_0 n_2}\ ,
\label{eq:Pcr}
\end{equation}
with $\lambda_0$ being the vacuum wave length, self-focusing occurs with the self-focusing angle and self-focusing distance (ignoring diffraction effects)
\begin{align}
\theta_{\textindex{sf}} &= \sqrt{\frac{2 n_2 I}{n_0}}\ , & z_{\textindex{sf}} &= \frac{w_0}{\theta_{\textindex{sf}}}\ .
\label{eq:thetasf}
\end{align}
Here, $w_0$ is the radius of the incident beam. If the beam power is equal to the critical power, the balance of self-focusing and diffraction effects results in self-trapping of the beam. It is important to note that self-focusing or self-trapping are specific consequences of the solution of the Maxwell equations in the Kerr medium \eqref{eq:Kerr1}. Since this solution maps algebraically onto the solution in the transformation medium \eqref{eq:kerr4}, the same effects will be seen in this new medium. In fact, by a suitable choice of transformations, self-interaction effects can be precisely engineered using the TO approach. At the same time it should be realized that the description of self-interaction effects based on intensity dependent refractive index profiles (such as Eqs.\ \eqref{eq:Pcr} and \eqref{eq:thetasf}) are not covariant and thus in general do not hold in the transformation medium. Indeed, while in the Kerr medium only a scalar linear and nonlinear response are present, the more complicated transformation medium \eqref{eq:kerr4} in general is based on anisotropic material parameters. Quantities like a scalar refraction index thus are not defined in these media.

\section{Examples of transformed Kerr media}
\label{sec:Examples}

Until now we have mainly developed the generic formalism of nonlinear TO. In this section we discuss some explicit examples. As mentioned in the introduction already, two different fields of applications of nonlinear TO exist:
\begin{enumerate}
	\item Nonlinear TO can be used in exactly the same way as linear TO with the sole difference that the starting point in this case is a nonlinear medium. An example of this type is discussed in Section \ref{sec:fieldcon}.
	\item Nonlinear media exhibit interesting self-interaction effects, which can be engineered by means of nonlinear TO. In contrast to the previous application, these devices rely on a productive combination of the self-interaction effects and the imposed coordinate transformation (see Section \ref{sec:KerrExample}). 
\end{enumerate}

\begin{figure*}[t]
\begin{tabular}{cc}
 \includegraphics[width=0.5\linewidth]{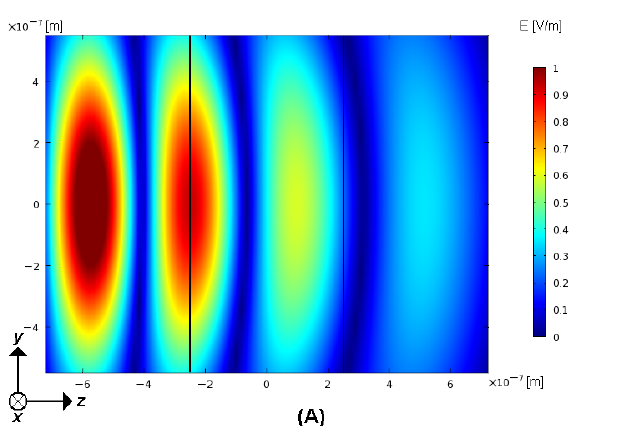} & \includegraphics[width=0.5\linewidth]{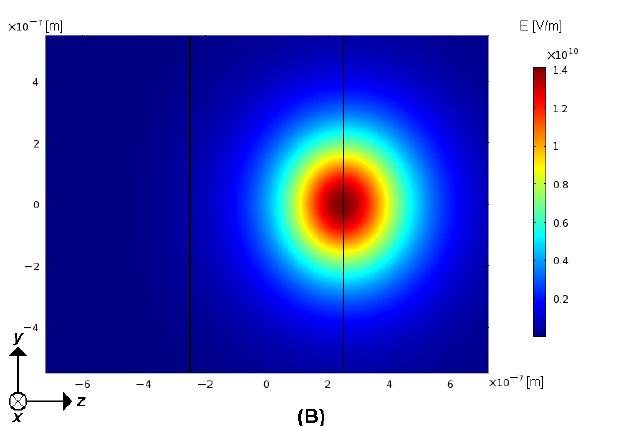}
\end{tabular}
\caption{Illustration of the laser beam in free space (A) and in a Kerr medium without transformation (B). Vertical lines indicate the region where the transformation will be applied in the second step, in this plot they are presented for illustrative purposes only.}
 \label{fig:free}
\end{figure*}\subsection{The nonlinear field concentrator}
\label{sec:fieldcon}
As mentioned in the introduction, many different applications of linear TO have been proposed. Most of these applications also work in nonlinear TO, with the important difference that in nonlinear TO the solutions in a Kerr medium are transformed as opposed to the free space solutions in standard linear TO. As a specific example we consider the field concentrator, whose transformations in cylindrical coordinates $(r,\phi,z)$ are \cite{Rahm:2008Cn}
\begin{gather}
 \label{ex01}
  \bar r =
  \begin{cases}
    \frac{R_1}{R_2} r\ , & 0<r<R_2\ ;\\
    \frac{R_3-R_1}{R_2-R_1} r - \frac{R_2-R_1}{R_3-R_2} R_3\ , & R_2<r<R_3\ ;\\
    r\ , & r>R_3\ ;
  \end{cases}\\
   \bar \phi = \phi\ ; \qquad \bar z = z\ ;
\end{gather}
with $0<R_1<R_2<R_3$. This transformation compresses a cylinder with radius $R_2$ in the original space into a cylinder with a smaller radius $R_1$, conversely space is stretched in the region $R_1 < \bar r < R_3$. In this way, the electromagnetic fields are concentrated in the central cylinder with radius $R_1$. Applying these transformation rules to the inverse metric in cylindrical coordinates, $[\gamma^{ij}] = \mbox{diag}(1,1/r^2,1)$, one obtains
\begin{equation}
 [\bar \gamma^{ij}] =
 \begin{cases}
    \mbox{diag}\left(\frac{R_1^2}{R_2^2}, \frac{R_1^2}{R_2^2} \frac{1}{\bar r^2}, 1\right)\ , & 0<\bar r<R_1\ ;\\
    \mbox{diag}\left(\frac{(R_3-R_1)^2}{(R_3-R_2)^2}, \frac{1}{\rho(\bar r)^2}, 1\right)\ , & R_1<\bar r<R_3\ ;\\
    \mbox{diag}(1,\frac{1}{\bar r^2}, 1)\ , & \bar r>R_3\ ;
 \end{cases}
\end{equation}
where
\begin{equation}
 \rho(\bar r) = r(\bar r)= \frac{R_3-R_2}{R_3-R_1} \bar r + \frac{R_2-R_1}{R_3-R_1} R_3\ .
\end{equation}
The determinants of the metrics are found as
\begin{align}
 \sqrt{\gamma} &= r\ , & \sqrt{\bar \gamma} &=
 \begin{cases}
   \frac{R_2^2}{R_1^2} \bar r\ , & 0<\bar r<R_1\ ;\\
   \frac{R_3-R_2}{R_3-R_1} \rho(\bar r)\ , & R_1<\bar r<R_3\ ;\\
    \bar r\ , & \bar r>R_3\ .
 \end{cases}
\end{align}
With this result the relative permittivity and permeability follow from Eq.\ \eqref{TOrel} as \cite{Rahm:2008Cn}
\begin{multline}
\label{ex04}
 [\tilde \varepsilon_R^{ij}] = [\tilde \mu_R^{ij}]=\\ =
 \begin{cases}
    \mbox{diag}\left(1, 1, \frac{R_2^2}{R_1^2}\right)\ , & 0<r<R_1\ ;\\
    \varrho \frac{\rho(r)}{r}\; \mbox{diag}\left(\frac{1}{\varrho^2}, \frac{1}{\rho(r)^2}, 1\right)\ , & R_1<r<R_3\ ;\\
    \mbox{diag}(1,\frac{1}{r^2}, 1)\ , & r>R_3\ .
 \end{cases}
\end{multline}
Here, we have used the abbreviation $\varrho = (R_3-R_2)/(R_3-R_1)$.
From this result the nonlinear field concentrator follows immediately. From Eq. \eqref{eq:Kerr2} the linear terms are given by the expression \eqref{ex04} -- up to a rescaling of permittivity by the relative permittivity of the original medium in Eq. \eqref{eq:Kerr2}. The transformation of the nonlinear term is slightly different, since this part is not rescaled by the determinants of the metrics. Thus, as the only missing piece
\begin{equation}
 \hat \chi \cdot [\bar \gamma^{ij}] = \hat \chi \cdot
  \begin{cases}
    \mbox{diag}\left(\frac{R_1^2}{R_2^2}, \frac{R_1^2}{R_2^2} \frac{1}{r^2}, 1\right)\ , & 0< r<R_1\ ;\\
    \mbox{diag}\left(\frac{(R_3-R_1)^2}{(R_3-R_2)^2}, \frac{1}{\rho(r)^2}, 1\right)\ , & R_1<r<R_3\ ;\\
    \mbox{diag}(1,\frac{1}{r^2}, 1)\ , & r>R_3\ ;
 \end{cases}
\end{equation}
is found as the relevant equation of the nonlinear field concentrator.

The nonlinear field concentrator compresses the electromagnetic field lines in a cylinder with radius $R_2$ into a cylinder with smaller radius $R_1$, conversely fields are stretched around this smaller cylinder. This working principle is completely independent of the specific electromagnetic fields in the material.

\subsection{Engineering self-focusing devices}
\label{sec:KerrExample}
The above example does not make use of the special behavior of nonlinear media. Similarly to linear TO devices, its main feature (namely, concentrating fields in the cylinder with radius $R_1$) applies independently of the specific form of the electromagnetic fields which are transformed. However, an important aspect of nonlinear media is their interesting response to specific electromagnetic field configurations. An interesting question thus arises, whether such effects also can be engineered using TO. In the second example we illustrate the power of TO in such an application as a self-focusing device. To keep the medium parameters at an acceptable level of complication, we require all medium tensors to be diagonal matrices in Cartesian coordinates, but still allow them to be anisotropic. For a pulse propagating in the $z$-direction an interesting class of transformations obeying this constraint is\footnote{More complicated manipulations are possible at the price of an inhomogeneous, nonlinear medium with the transformation 
\begin{align*}
 \bar z &= f(z)\ ,&(\bar \gamma^{ij}) &= \mbox{diag}\left( 1, 1, (f')^2\right) \ , & \sqrt{\bar \gamma} &= 1/f' \ ,
\label{eq:ex4}
\end{align*}
where the function $f(\bar z)$ should be monotonically increasing.}
\begin{equation}
 \bar z = \frac{1}{\alpha} z\ , \qquad \alpha>0\ .
\label{eq:ex1}
\end{equation}
If the original coordinate system $(x,y,z)$ is assumed to be Cartesian, then the above transformation results in
\begin{align}
  [\bar \gamma^{ij}] &= \begin{pmatrix} 1 & 0 & 0 \\ 0&1&0 \\ 0&0& \frac{1}{\alpha^2} \end{pmatrix}\ , & \sqrt{\bar \gamma} &= \alpha \ ,
\end{align}
and thus we find the transformed medium parameters as
\begin{align}
\tilde D^i &=  \left( \alpha \varepsilon + \hat \chi \tilde{\bm D} \cdot \tilde{\bm E} \right) \bar \gamma^{ij} \tilde{E}_j\ ,  & \tilde B^i &= \alpha \mu_0  \bar \gamma^{ij} \tilde H_j\ .
\label{eq:ex2}
\end{align}
The parameter $\alpha$ defines how the typical length of self-interaction effects (e.g.\ the self-focusing length) is shortened ($\alpha > 1$) or streched ($\alpha < 1$) with the transformation. If we wish to express the nonlinearity purely in terms of $E_i$, we can make use of the relation
\begin{multline}
 \tilde{\bm D} \cdot \tilde{\bm E} = \tilde{D}^i \tilde{E}_i = \alpha \varepsilon  \sum_{n=1}^{\infty} \hat\chi^{n-1} (\tilde{E}_i \bar \gamma^{ij} \tilde{ E}_j)^{n}\\ = \alpha \varepsilon \sum_{n=1}^{\infty} \hat\chi^{n-1} (E_x^2 + E_y^2 + \frac{1}{\alpha^2} E_z^2)^{n}\ .
\label{eq:ex3}
\end{multline}

\begin{figure*}[t]
\begin{tabular}{cc}
 \includegraphics[width=0.5\linewidth]{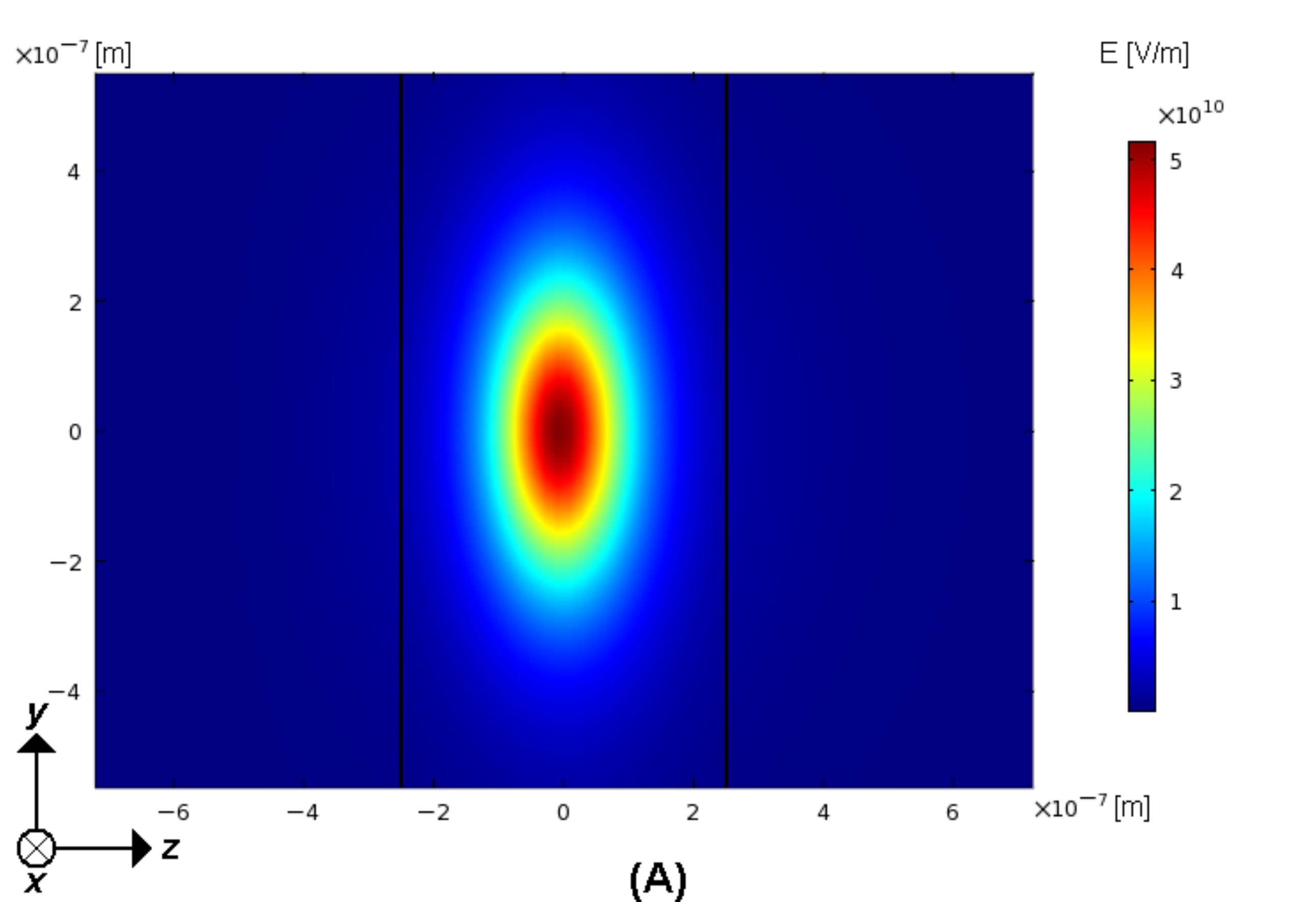} & \includegraphics[width=0.5\linewidth]{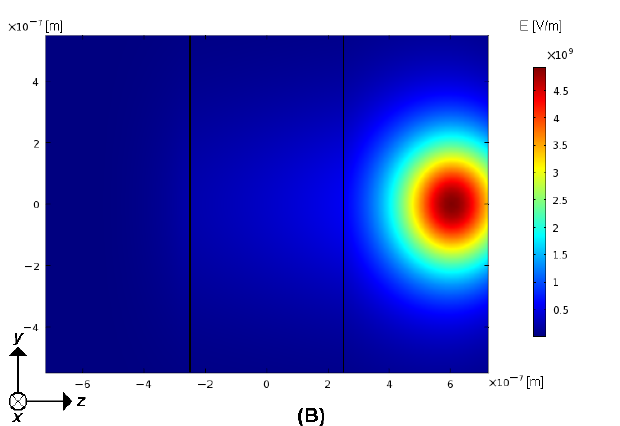}
\end{tabular}
\caption{The effect of transformations with $\alpha=2$ (A) and $\alpha=0.25$ (B). The transformation is applied to the central region indicated by the two vertical lines only, the left and right parts remain unchanged compared to the Kerr medium in Fig.~\ref{fig:free}.}
 \label{fig:alpha}
\end{figure*}
\begin{figure*}[t]
\begin{tabular}{cc}
 \includegraphics[width=0.5\linewidth]{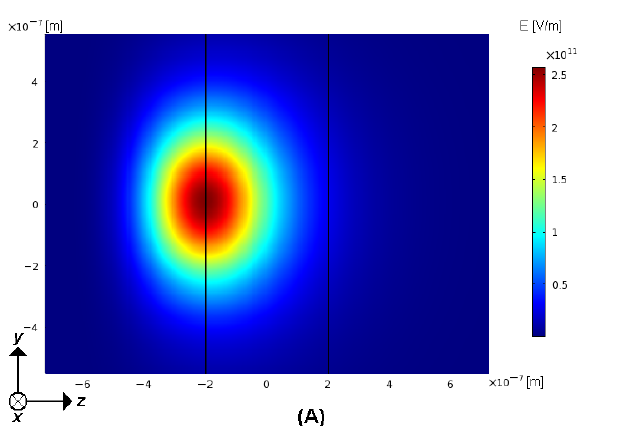} & \includegraphics[width=0.5\linewidth]{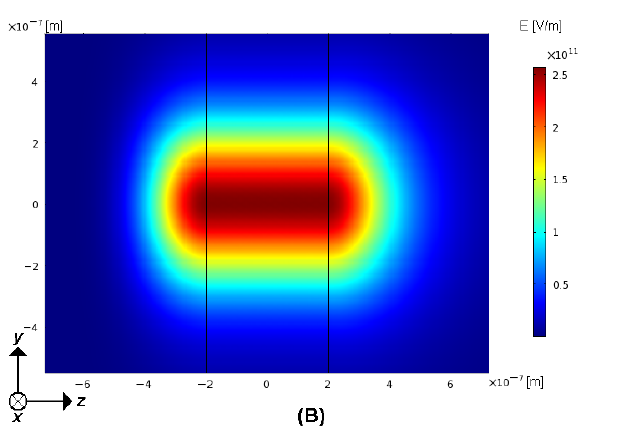}
\end{tabular}
\caption{The untransformed Kerr medium with $\chi=1.9\cdot 10^{-9}\; \frac{\mbox{Cm}}{\mbox{V}^3}$ (A). On the right hand side (B), $\alpha=0.01$ is chosen in the central region indicated by the two vertical lines, such that the focal point freezes and builds a concentrated flux tube.}
 \label{fig:frozen}
\end{figure*}

Next, using a time-domain simulation in COMSOL Multiphysics \cite{COMSOL}, the possibility of engineering the self-focusing distance is demonstrated numerically. The results are shown in Figs.\ \ref{fig:free}--\ref{fig:frozen}. Some explanations are in order: The left boundary of the figures is the excitation boundary, where a ``laser beam'' is modeled by the Gaussian beam
\begin{equation}
 E(t)=E_0 \exp\left(-\frac{y^2}{l^2}\right) \sin(2\pi \nu t)\ ,
\end{equation}
with the choice $E_0 = 1\; \mbox{V}/\mbox{m}$, $l=3\cdot 10^{-13}\;\mbox{m}$ and $\nu=300\;\mbox{THz}$. The electric field points in the $x$-direction and propagates in the positive $z$-direction, i.e.\ from left to right in all figures. The side length of the plotted area is $1.5\cdot 10^{-6}\;\mbox{m}$ in $y$ and $2\cdot 10^{-6}\;\mbox{m}$ in the $z$-direction, with the snapshot taken at $t=10^{-14}\;\mbox{s}$; these numbers have been chosen for convenience and do not exhibit a deeper physical interpretation. The propagation of the beam in free space and in a Kerr medium
\begin{align}
D^i &= \left( \varepsilon_0 + \chi \bm E^2\right) \gamma^{ij} E_j\ , & \chi &= 1.5\cdot 10^{-10}\; \frac{\mbox{Cm}}{\mbox{V}^3}\ ,
\label{explicitKerr}\\
 B^i &= \mu_0 \gamma^{ij} H_j\ , &&
\end{align}
is illustrated in Fig.\ \ref{fig:free}. This Kerr medium is used as a starting point of our manipulations; in the central region of the simulation area, indicated by the two vertical lines, transformations with various values of $\alpha$ are chosen, while in the left and right areas all medium parameters are kept fixed. Results for $\alpha=0.25$ and $\alpha=2$ are shown in Fig.\ \ref{fig:alpha}. Since COMSOL Multiphysics does not allow for implicit definitions of the constitutive equation, the expansion \eqref{eq:ex3} was used for the Kerr nonlinearity, whereby all the terms except for the leading one have been dropped. Despite this approximation, it is clearly seen that the self-focusing length is shortened for $\alpha > 1$ and elongated for $\alpha<1$. Furthermore, the interfaces between the untransformed and the transformed Kerr media are reflectionless; it can be shown on general grounds \cite{Bergamin:2009In} that this must be true for any transformation of type \eqref{eq:ex1} at interfaces defined by a constant coordinate $z$. If the self-focusing point lies in the transformed region (as is the case for $\alpha=2$), the self-focusing pattern is deformed by the anisotropy of the medium parameters \eqref{eq:ex2} (space is squeezed in the $z$-direction but not in the $x$ and $y$-directions).

Completely new effects can be obtained by choosing extreme medium parameters, for example a very small value of $\alpha$ can yield a freezing of the focal points. To demonstrate this we change the untransformed Kerr medium in such a way that the focal point is located exactly at the interface of the left and center areas (see Fig.\ \ref{fig:frozen}). In the second step a transformation is again applied to the central region, in this example with a very small value of $\alpha=0.01$. In this way the width of the central region becomes very small in the transformed space, the self-interaction effects can hardly evolve and consequently the self-focusing is frozen. As a result we obtain an almost ideal flux tube over the whole area of the transformation medium.

This example shows that genuine characteristics of nonlinear media can be engineered by means of nonlinear TO. Of course, such devices will only work for a class of electromagnetic fields that exhibit a certain characteristic, in contrast to the examples of the previous subsection, whose workability is completely general. Still, this example shows a completely new field of applications for nonlinear TO, which does not exist in conventional linear TO.

\section{Conclusions}
\label{sec:Conclusions}
In this paper we have developed the formalism of nonlinear transformation optics. Standard transformation media, which can be interpreted as mimicking empty space, are inherently linear. Therefore, in nonlinear TO the original space, on which the transformation is applied, has to be nonlinear. In a generic expansion of the constitutive relations in terms of $\bm E$ and $\bm H$ the formalism extends straightforwardly from the linear case, at least for purely spatial transformations. Less immediate are situations such as the Kerr nonlinearity, where the nonlinear contributions are expanded in terms of $\bm E^2$ and $\bm H^2$, since these quantities implicitly depend on the metric: $\bm E^2 = E_i \gamma^{ij} E_j$. We have shown that this problem can be circumvented by defining implicit constitutive relations in terms the electric and magnetic energy densities, $\bm D \cdot \bm E$ and $\bm B \cdot \bm H$. In this formulation nonlinear TO retains all the intuitive design properties of standard linear TO, such that the (nonlinear) field configurations simply can be considered as sticked to the coordinate lines during the transformation. 

Most applications of linear TO are insensitive to the exact form of the electromagnetic fields, in other words, their functional principle works -- at least theoretically -- for any field configuration. Such devices extend straightforwardly to nonlinear TO, which was demonstrated in an example of an electromagnetic concentrator in a Kerr medium. Besides such applications, nonlinear TO allows for a completely new type of devices: Since nonlinear media exhibit interesting self-interaction effects, which depend on the exact field configuration, TO can be used to engineer such effects. As a specific example we have shown that the self-focusing effect of a laser beam in a Kerr medium can be engineered by means of nonlinear TO. Although the working principle of such devices is restricted to exactly those field configurations that exhibit the desired self-interaction effect in the original Kerr medium, the design principles of TO are retained: By a suitable compression or stretching of space the self-focusing length can be altered; if space is stretched by a large factor exactly at the focusing point, the focusing area is ``frozen'', which results in a high intensity flux tube.

The phenomenology of nonlinear media is much richer than we could treat in this article. It would be interesting to study the implications of nonlinear TO in more detail, including such aspects as the global solutions (solitons), second harmonic generation or different self-interaction effects. On a different route of research it should be investigated how the proposed nonlinear materials could be realized as artificial media (metamaterials).

\begin{acknowledgments}
This work was partially funded by the Academy of Finland and Nokia through the Center-of-Excellence program. L.~Bergamin was partially supported by the Academy of Finland, project no.\ 124204. The work of P.~Alitalo was supported by the Academy of Finland through post-doctoral project funding.
\end{acknowledgments}

%

\end{document}